\documentclass[twocolumn,showpacs,preprintnumbers,amsmath,amssymb]{revtex4}
\usepackage{graphicx}
\usepackage{dcolumn}
\usepackage{bm}
\usepackage{color}

\begin{document}
\preprint{APS/123-QED}

\title{Birth, Death and Flight: A Theory of 
Malthusian Flocks}
\author{John Toner}
\affiliation{Department of Physics and Institute of Theoretical
Science, University of Oregon, Eugene, OR 97403}
\date{\today}
\begin{abstract}
I study ``Malthusian Flocks": moving aggregates of self-propelled entities (e.g., organisms, cytoskeletal actin, microtubules in mitotic spindles) that reproduce and die. Long-ranged order (i.e., the existence of  a non-zero average velocity $\left<\vec v (\vec r, t) \right>\ne \vec 0$) is possible in these systems, even in  spatial dimension $d=2$. Their spatiotemporal scaling structure can be determined exactly in $d=2$;  furthermore, they  lack both  the longitudinal sound waves and the giant number fluctuations found in immortal flocks. Number fluctuations are very {\it persistent}, and propagate along the direction of flock motion, but at a different speed.\end{abstract} 
\maketitle

\vskip.5cm 

Flocking \cite{boids}  -- the coherent motion of large numbers of
organisms -- 
spans a wide range of length scales:  from
kilometers (herds of wildebeest) to microns (microorganisms
\cite{dictyo,rappel1}; mobile  macromolecules in living cells\cite{actin,microtub}).
It is also \cite{Vicsek}  a dynamical version of ferromagnetic ordering. A``hydrodynamic" theory of flocking \cite{TT} shows 
that, unlike equilibrium ferromagnets \cite{MW},  
flocks {\it can}
spontaneously break a continuous
symmetry (rotation
invariance) by developing long-ranged order, (i.e., a non-zero average velocity $\left<\vec v (\vec r, t) \right>\ne \vec 0$) in spatial dimensions $d=2$, 
even with only short
ranged interactions. 

Many quantitative predictions  of the hydrodynamic theory , including the stability of long-ranged order in $d=2$, the existence of propagating, dispersion-less  sound modes with non-trivial direction dependence of their speeds, and the presence of anomalously large number fluctuations, 
agree with numerical simulations\cite{TT,Chate1}, and  experiments on self-propelled molecules\cite{Rao}.

However, a recent re-analysis\cite{me recent} of this hydrodynamic theory has 
cast doubt on the claim that exact scaling exponents could be determined for flocks in $d=2$. This is due to the erroneous neglect in\cite{TT} of non-linearities arising from the local number density dependences of various phenomenological parameters; these  non-linearities could 
change the scaling exponents from those claimed in \cite{TT}. 

In this paper, I show that these difficulties can be avoided
in flocks {\it without} number conservation. A number of real systems\cite{SR, Lev} lack number conservation, including growing bacteria colonies\cite{Lev},   and ``treadmilling" 
molecular motor propelled biological macromolecules in a variety of intra-cellular structures, including the cytoskleton, and mitotic spindles\cite{actin, microtub}, in which molecules 
are being created and destroyed as they move. Hence, he study of such systems is not only convenient, but experimentally relevant. The most obvious example of flocking - namely, actual birds - is clearly {\it not} a good example of a Malthusian flock.
I will henceforth use the 
term ``boid"  \cite{boids}  to refer to the moving, self-propelled entities that make up the flock. 

I will treat systems with the same symmetries as were considered in earlier work on immortal flocks\cite{TT}: {\it orientationally ordered, translationally disordered}  phases (i.e., phases  with $\left<\vec v (\vec r, t) \right>\ne \vec 0$ that are uniform in space and time) in systems with short-ranged, rotation invariant interactions, moving through (or on, in $d=2$) a {\it fixed} background medium that breaks Galilean invariance. 

Since my treatment is hydrodynamic, it only describes large systems at long length and time scales. However, it becomes asymptotically exact in that limit.

The removal of number  conservation leads to profound changes. The sound modes of number conserving (hereafter ``immortal") flocks disappear, and are replaced by longitudinal velocity fluctuations which  drift in the direction of flock motion with a speed $\gamma\ne v_0$, where $v_0$ is the mean speed of the flock. 
While drifting, these modes also spread diffusively  along the direction of flock motion, and hyperdiffusively perpendicular to that direction. 

In both Malthusian and immortal flocks, anomalous hydrodynamics stabilizes long-ranged orientational order (i.e., $\left<\vec v (\vec r, t) \right>\ne \vec 0$)  in spatial dimension $d=2$.
In  Malthusian flocks, the order  outlives the boids: the persistence time diverges as the number of boids 
$N\rightarrow\infty$, while the lifetime of the boids  remains finite in this limit.

The scaling exponents of  the hydrodynamics can be determined {\it exactly} in  spatial dimension $d=2$ for Malthusian flocks, while, as discussed above,  those in immortal flocks cannot.
These exponents are: the {\it dynamical} exponent
$z$ for the scaling of  timescales $t(L_\perp)\propto L_\perp^z$ with  length scale  $ L_\perp$ perpendicular to the direction of flock motion; an {\it anisotropy} exponent $\zeta$ for the scaling of distances $L_\parallel(L_\perp)\propto L_\perp^\zeta$ parallel to the direction of motion with $ L_\perp$ ; and a  ``roughness" exponent $\chi$ relating the scale of velocity fluctuations to $L_\perp$ via
$\delta v\propto L_\perp^\chi$. I find that, for Malthusian flocks in spatial dimension $d=2$, 
these exponents are:

\begin{eqnarray} 
\zeta = {3\over 5}~~,~~~
z = {6\over 5}~~,~~~
\chi = -{1\over 5} ~~.
\label{exact exp3} 
\end{eqnarray}

The velocity field can have  long-ranged order; ($\left<\vec v (\vec r, t) \right>\ne \vec 0$) in $d=2$, because
 the roughness exponent $\chi(d=2)<0$.

Number fluctuations in Malthusian flocks exhibit anomalous {\it persistence}:
the experimentally observable density-density correlation function 
\begin{eqnarray} 
C_\rho (\vec r , t) \equiv \left<\delta \rho(\vec{r}\,', t')\delta \rho(\vec{r}\,'+\vec{r}, t'+ t)\right>  ~,\label{Crho} 
\end{eqnarray} 
where $\delta \rho(\vec{r}, t)\equiv \rho(\vec{r}, t) - \rho_0$ is the departure of the local number density of boids  $ \rho(\vec{r}, t)$ from its mean value $ \rho_0$,
decays algebraically with time 
at a fixed point in space:
\begin{eqnarray} 
C_\rho (\vec r = \vec{0},  t)  \propto |t|^{-4} ~,\label{persist} 
\end{eqnarray} 
in spatial dimensions $d=2$, while for a point translating along the direction $\hat x_{\parallel}$ of flock motion at the ``drift" speed $\gamma$  the decay is 
even slower: in both $2$ and $3$ spatial dimensions, I find
\begin{eqnarray} 
C_\rho (\vec r = \gamma t  \hat x_{\parallel},  t)\propto |t|^{-2} ~.\label{sweet spot} 
\end{eqnarray} 
These should be contrasted with the exponential decay  with time of density fluctuations that occurs in a {\it disordered} Malthusian flock (i.e., one in which $<\vec{v}> = \vec{0}$).

Since $C_\rho$ and $C_v$ can be constructed from {\it any} set of high-resolution images of a moving flock, as has been done for flocks of starlings in reference\cite{Andrea and Irene} and for a related correlation function of bacteria in\cite{Lev}, it should not be difficult to test these predictions.

I will now outline the derivation of these results.

My starting equation of motion for the velocity is exactly that of an 
``immortal" flock\cite{P2}: 
\begin{eqnarray}
&&\partial_{t}
\vec{v}+\lambda_1 (\vec{v}\cdot\vec{\nabla})\vec{v}+
\lambda_2 (\vec{\nabla}\cdot\vec{v})\vec{v}
+\lambda_3\vec{\nabla}(|\vec{v}|^2)
 =\nonumber \\&&
\alpha\vec{v}-\beta
|\vec{v}|^{2}\vec{v} -\vec{\nabla} P_1 -\vec{v} 
\left( \vec{v} \cdot \vec{\nabla}  P_2 (\rho,|\vec{v}|) \right)\nonumber \\&&+D^o_{B} \vec{\nabla}
(\vec{\nabla}
\cdot \vec{v}) + D_{T}\nabla^{2}\vec{v} +
D_{2}(\vec{v}\cdot\vec{\nabla})^{2}\vec{v}+\vec{f}
\label{vEOM}
\end{eqnarray}
where all of the parameters $\lambda_i (i = 1 \to 3)$,
$\alpha$, $\beta$,$D^o_B$, $D_{T,2}$ and the  ``pressures'' $P_{1,2}(\rho,
|\vec{v}|)$ are, in general, functions of the boid number density $\rho$ and the magnitude
$|\vec{v}|$ of the local velocity. 
I will expand  $P_{1,2}(\rho,
|\vec{v}|)$ about $\rho_0$: 
$P_i(\rho) = P_i^0 + \sum_{n=1}^{\infty}\sigma_{i,n}(|\vec{v}|)\delta\rho^n$, where $i=1,2$.

In (\ref{vEOM}), $\beta$, $D^o_{B}$, $D_{2}$ and $D_{T}$ are all
positive, while
$\alpha < 0$ in the disordered phase and $\alpha>0$ in
the ordered state.

The
$\alpha$ and
$\beta$ terms give
$\vec{v}$ a nonzero magnitude $v_0=\sqrt{{\alpha} \over
{\beta}}$   
in the ordered phase.
The diffusion constants $D_{B,T,2}$  reflect the tendency of  ``boids" to follow their neighbors.
The $\vec{f}$ term is a random Gaussian white noise, mimicking  errors made by the boids, with correlations:
\begin{eqnarray}
   <f_{i}(\vec{r},t)f_{j}(\vec{r'},t')>=\Delta
\delta_{ij}\delta^{d}(\vec{r}-\vec{r'})\delta(t-t')
\label{white noise}
\end{eqnarray}
where $\Delta ={\rm constant}$, and $i , j$ label
vector components. 
The ``anisotropic pressure'' $P_2(\rho, |\vec{v}|)$ in
(\ref{vEOM}) is only allowed due to the non-equilibrium nature of the
flock; in an equilibrium fluid such a term is forbidden by Pascal's
Law.  In earlier work \cite{TT} this term was ignored. 

Note that  (\ref{vEOM}) is {\it not}
Galilean invariant;  it holds only in the frame of the fixed medium through or on which  the creatures move.
 
I now need an equation of motion for $\rho$. In immortal flocks, this is just  the usual continuity equation of compressible fluid dynamics. For Malthusian flocks, it must also include the effects of birth and 
death. I will assume that the death rate goes up as the density goes up 
(the ``Malthusian" assumption\cite{Malthus}), so that the difference between the two - that 
is, the net, local growth rate of number density in the absence of motion, 
which I'll call $g(\rho)$ - vanishes at some fixed point density $\rho_0$, 
with larger densities decreasing (i.e., $g(\rho > \rho_0) < 0)$, and smaller 
densities increasing (i.e., $g(\rho < \rho_0) > 0)$. 

The equation of motion for the density is now simply: 
\begin{eqnarray}
\partial_t\rho +\nabla\cdot(\vec{v}\rho)=g(\rho)~~.
\label{conservation}
\end{eqnarray} 
Note that in the absence of birth and death, $g(\rho) = 0$, and equation 
(\ref{conservation}) reduces to the usual continuity equation, as it should, since 
``boid number" is then conserved.

Since birth and death quickly restore the 
fixed point density $\rho_0$, I will 
write $\rho(\vec r, t) 
= \rho_0 + \delta\rho(\vec r, t)$ and expand both sides of equation 
(\ref{conservation}) to leading  order in $\delta\rho$. This gives 
$\rho_0\vec \bigtriangledown \cdot \vec v \cong g' (\rho_0)\delta\rho , 
$ where I've dropped the $\partial_t\rho$ term relative to the $ g' (\rho_0)\delta\rho$ term since I'm interested in the hydrodynamic limit, in which the fields evolve extremely slowly.
This equation can be readily solved to give 
\begin{eqnarray} 
\delta\rho \cong {\rho_0 \vec{\nabla} \cdot \vec v \over g' 
(\rho_0)} \equiv - {\Delta D_B^{(1)}\over \sigma_{1, 1}}(\vec{\nabla}
\cdot \vec v) \label{rho-v} 
\end{eqnarray} 
where $\Delta D_B^{(1)}$ is a positive constant,  and $\sigma_{1, 1}$ is  the first expansion coefficient  for $P_1$.
I can now insert this solution (\ref{rho-v}) for $\delta\rho$ in terms of $\vec 
v$ into the isotropic pressure $P_1$; the resulting equation of motion  for $\vec v$ is:
\begin{eqnarray}
&&\partial_{t}
\vec{v}+\lambda_1 (\vec{v}\cdot\vec{\nabla})\vec{v}+
\lambda_2 (\vec{\nabla}\cdot\vec{v})\vec{v}
+\lambda_3\vec{\nabla}(|\vec{v}|^2)
 =\nonumber \\&&
\alpha\vec{v}-\beta
|\vec{v}|^{2}\vec{v} -\vec{v} 
\left( \vec{v} \cdot \vec{\nabla}  P_2 (\rho,|\vec{v}|) \right)\nonumber \\&&+D^{(1)}_{B} \vec{\nabla}
(\vec{\nabla}
\cdot \vec{v}) + D_{T}\nabla^{2}\vec{v} +
D_{2}(\vec{v}\cdot\vec{\nabla})^{2}\vec{v}+\vec{f}~,
\label{vEOM2}
\end{eqnarray}
where I've defined $D^{(1)}_B\equiv D^o_B+\Delta D_B^{(1)}$.
Taking the dot product of both sides of (\ref{vEOM2}) with $\vec{v}$ itself,
and defining $U(|\vec{v}|)\equiv\alpha(|\vec{v}|, \rho)-\beta(|\vec{v}|, \rho)|\vec{v}|^{2}$, I obtain:
\begin{widetext}
\begin{eqnarray}
{1\over 2}\left(\partial_{t}|\vec{v}|^2+(\lambda_1+2\lambda_3)(\vec{v}\cdot\vec{\nabla})|\vec{v}|^2\right) + \lambda_2(\vec{\nabla}\cdot\vec{v})|\vec{v}|^2&=& U(|\vec{v}|)|\vec{v}|^{2}-|\vec{v}|^{2}\vec{v} \cdot \vec{\nabla}  P_2 +D^{(1)} _B\vec{v}\cdot\vec{\nabla}
(\vec{\nabla}\cdot \vec{v}) 
\nonumber \\&+& 
D_{T}\vec{v}\cdot\nabla^{2}\vec{v} +D_{2}\vec{v}\cdot\left((\vec{v}\cdot\vec{\nabla})^{2}\vec{v}\right)+\vec{v}\cdot\vec{f}
\label{v parallel elim}~,
\end{eqnarray}
\end{widetext}

In the ordered state (i.e.,  in which $\left<\vec v (\vec r, t) \right>= v_0 \hat x_{\parallel}$), and 
I can 
expand the $\vec{v}$ equation of motion for small departures $ \delta \vec v (\vec r,t) $ of $\vec{v}(\vec r,t) $ from uniform motion with speed $v_0$:
\begin{eqnarray} 
\vec v (\vec r, t) = (v_0+\delta v_\parallel) \hat x_{\parallel} + \vec v_{\perp}(\vec r,t) ~~,
\label{6} 
\end{eqnarray} 
where, henceforth  $\parallel$ and  $\perp$ denote components along and perpendicular to the mean velocity, respectively.

In this hydrodynamic approach, 
I'm interested only in fluctuations $\vec{\delta v}(\vec{r}, t)\equiv\delta v_\parallel \hat x_{\parallel} + \vec v_{\perp}(\vec r,t)$ and $\delta \rho(\vec{r}, t)$
that vary slowly in space and time. 
Hence, terms involving spatiotemporal derivatives of 
$\vec{\delta v}(\vec{r}, t)$ and $\delta \rho(\vec{r}, t)$
are always negligible, in the hydrodynamic limit, compared to terms involving the same number of powers of fields with fewer spatiotemporal derivatives. Furthermore, the fluctuations 
$\vec{\delta v}(\vec{r}, t)$ and $\delta \rho(\vec{r}, t)$ can themselves be shown to be small in the long-wavelength limit. Hence, we need only keep terms in [\ref{v parallel elim}] up to linear order in 
$\vec{\delta v}(\vec{r}, t)$ and $\delta \rho(\vec{r}, t)$. The 
$\vec{v}\cdot\vec{f}$ term can likewise be dropped.

These observations can be used to eliminate many  terms in equation [\ref{v parallel elim}], and solve for the quantity $U \equiv ( \alpha(\rho , |\vec{v}|)-\beta (\rho ,
|\vec{v}|)|\vec{v}|^2)$; I obtain:
$U=\lambda_2 \vec{\nabla}\cdot\vec{v}
+\vec{v}\cdot\vec{\nabla}P_2$.
 Inserting this expression for $U$ back into equation [\ref{vEOM2}],
I find that $P_2$ and $\lambda_2$ cancel out of the $\vec{v}$ equation of motion, leaving, ignoring irrelevant terms:
\begin{eqnarray}&&\partial_{t}
\vec{v}+\lambda_1(\vec{v}\cdot\vec{\nabla})\vec{v}+\lambda_3 \vec{\nabla}(|\vec{v}|^2) =D_T\nabla^2 \vec v
\nonumber \\&&
+D^{(1)}_{B}  \vec \nabla(\vec \nabla\cdot \vec 
v) +
D_{2}(\vec{v}\cdot\vec{\nabla})^{2}\vec{v}
 + \vec f,
\label{vperpEOM} 
\end{eqnarray} 
This can be made into an equation of motion for $\vec{v}_\perp$ involving only $\vec{v}_\perp(\vec{r}, t)$ 
itself by projecting perpendicular to the direction of mean flock motion $\hat{x}_\parallel$, and eliminating $\delta v_\parallel$ using $U=\lambda_2 \vec{\nabla}\cdot\vec{v}
+\vec{v}\cdot\vec{\nabla}P_2$ and 
the expansion $
U\approx-\Gamma_1\delta v_\parallel - \Gamma_2 \delta \rho$, 
where 
I've defined $
\Gamma_1 \equiv -\left({\partial U
 \over \partial |\vec{v}|}\right)^0_{\rho}$ and 
$\Gamma_2 \equiv - \left({\partial U
 \over \partial \rho}\right)^0_{|\vec{v}|}$,
with
super-
or sub-scripts
$0$ denoting functions of  $\rho$ and 
$|\vec{v}|$ evaluated at $\rho = \rho_0$ and $ |\vec{v}|=v_0$. Doing this, and using (\ref{rho-v}) for $\rho$, I obtain:
\begin{eqnarray}&&\partial_{t}
\vec{v}_{\perp}+\gamma\partial_\parallel \vec{v}_\perp+\lambda_1 (\vec{v}_{\perp}\cdot\vec{\nabla}_{\perp})\vec{v}_{\perp} +\lambda_3 \vec \nabla_{\perp}\left(|\vec{v}_\perp|^2\right)=
\nonumber \\&&D_T\nabla_{\perp}^2 \vec v_{\perp} 
+D_{B}  \vec \nabla_{\perp}(\vec \nabla_{\perp}\cdot \vec 
v_{\perp}) + D_\parallel\partial_{\parallel}^{2}\vec v_{\perp} + \vec f_{\perp},
\label{vperpEOM} 
\end{eqnarray} 
where  I've defined $\gamma\equiv\lambda_1 v_0$, 
$D_B\equiv D^0_B+2v_0\lambda_3(\lambda_2-\Gamma_2\Delta D_B^{(1)} /\sigma_1)
/\Gamma_1$  
and $D_{\parallel} \equiv
D^0_{T}+D^0_{2}v_0^2$.
Changing co-ordinates to a new Galilean frame $\vec{r}\,'$ moving with respect to our original frame 
in the direction of mean flock motion at speed $\gamma$ - i.e., $ 
\vec{r}\,'\equiv\vec{r}-\gamma t  \hat x_{\parallel}
$ -
gives
\begin{eqnarray}\partial_{t}
\vec{v}_{\perp}+\lambda_1 (\vec{v}_{\perp}\cdot\vec{\nabla}_{\perp})\vec{v}_{\perp}=&&D_T\nabla_{\perp}^2 \vec v_{\perp} 
+D_{B}  \vec \nabla_{\perp}(\vec \nabla_{\perp}\cdot \vec 
v_{\perp}) \nonumber \\&&+D_\parallel\partial_{\parallel}^{'2}\vec v_{\perp}  + \vec f_{\perp}~~.
\label{vperpEOM'} 
\end{eqnarray}

Ignoring the non-linear term $\lambda_1$ in this equation of motion
gives a noisy, anisotropic, vectorial diffusion equation. This can be readily solved for the mode structure and fluctuations by spatiotemporal Fourier transformation, and has $d-1$ diffusing modes
in spatial 
dimension $d$.  These separate into $d-2$ ``transverse" modes (i.e., modes with $\vec{v}_\perp$ perpendicular to  $\vec{q}_\perp$, all with the same imaginary eigenfrequency:
$
-i \omega_{_T} = D_{T}|\vec {q}_{_\perp}|^2 + D_{\parallel}q_\parallel^2
$.
The remaining diffusive mode  (the {\it only} mode in  $d=2$) is  ``longitudinal"  (i.e., has $\vec{v}_{_\perp}$ {\it along}   $\vec{q}_{_\perp}$), with frequency
$
-i \omega_{_L} = D_{_\perp}|\vec {q}_{_\perp}|^2 + D_{\parallel}q_\parallel^2
$,
where $D_{_\perp}\equiv D_B + D_T$.


Because the dynamics described above is in the Galileanly boosted frame,
the dynamics in the original reference frame $\vec{r}$ will have a steady drift at velocity $\gamma$ superposed on the diffusive motion described above; that is, both eigenfrequencies get $\gamma q_\parallel$ added to them.

I can also calculate  the real-space  velocity fluctuations $\left<|\vec {v}_\perp (\vec r, t)|^2 \right> $ in this linearized approximation; I find that, {\it in this  approximation}, this diverges in all   $d\le2$. This is  analogous to the Mermin-Wagner theorem\cite{MW}  in equilibrium magnets.
However, as in 
immortal flocks\cite{TT}, this ``Mermin-Wagner" result, and all the linearized scaling laws, are invalidated for 
$d\le 4$ by 
the $\lambda_1$ term in (\ref{vperpEOM'}).

To show this here, I'll analyze equation (\ref{vperpEOM'}) using the dynamical Renormalization Group(RG)\cite{FNS}. 

The dynamical RG starts by averaging the equations of motion over the short-wavelength fluctuations: i.e.,   those with support in the ``shell" of Fourier space $b^{-1} \Lambda \le |\vec{q}| \le \Lambda$, where $\Lambda$ is an ``ultra-violet cutoff", and $b$ is an arbitrary rescaling factor. Then, one  rescales lengths, time,  and $\vec{v}_{_\perp}$ in equation (\ref{vperpEOM'}) according to 
$\vec{v}_{_\perp} = b^\chi \vec{v}_{_\perp}^{\,\prime}$, $\vec{r}_\perp=b\vec{r}_\perp^{\,\prime}$, $r'_
\parallel=b^{\zeta}(r'_\parallel)'$, and $t = b^zt'$ to restore the ultra-violet cutoff  to $\Lambda$.  
This leads to a new equation of motion of the same form as (\ref{vperpEOM'}), but with 
``renormalized" values (denoted by primes below) of the parameters given by:
\begin{eqnarray} 
D'_{B, T} = b^{z-2}(D_{B,T}+  {\rm 
graphs} )~~, 
\label{rescale Dperp}    
\end{eqnarray}
\begin{eqnarray} 
D'_\parallel=b^{z-2\zeta}(D_\parallel + {\rm 
graphs} )~~,
\label{rescale Dpar}   
\end{eqnarray}
\begin{eqnarray} 
\Delta' = b^{z-\zeta-2\chi+1-d}(\Delta+{\rm 
graphs} )~,
\label{rescale Delta}    
\end{eqnarray}
\begin{eqnarray} 
\lambda_{1,3}' = b^{z+\chi-1}(\lambda_{1,3}+{\rm 
graphs} )~~
,
\label{rescale lambda} 
\end{eqnarray} 
where ``graphs" denotes contributions from integrating out the short wavelength degrees of freedom.
If we ignore these graphical corrections (valid for $\lambda_{1,3}$ small), and choose
$z$, $\zeta$, and $\chi$ to keep the linear parameters $D_{B,T,\parallel}$ and $\Delta$ fixed, equation (\ref{rescale lambda})  implies that an initially  small  $\lambda_1$ will grow for all 
$d\le 4$, meaning the linearized theory 

It is possible to  get {\it exact} exponents in $d=2$. This is  because the nonlinearities - the $\lambda_1$ and $\lambda_3$ terms - 
in (\ref{vperpEOM'}) add up to a total derivative in $d=2$ (specifically,  $\left({\lambda_1\over 2}+\lambda_3\right)\partial_\perp v_\perp^2$), since the  $\perp$ subspace is one dimensional in $d=2$. 
In contrast, in immortal flocks, $v$-$\rho$ non-linearities arising from the $\rho$-dependence of $\lambda_1$
cannot be written as total derivatives, making it impossible to obtain exact exponents, a fact missed by\,\cite{TT}.
Here, because   the $\lambda_1$ term {\it is} a total $\perp$-derivative, it can only graphically renormalize terms  involving $\perp$-derivatives themselves. Hence,  the graphical corrections to  
$D_{\parallel}$ and $\Delta$ in equations (\ref{rescale Dpar}) and (\ref{rescale Delta}) vanish. Hence, at a fixed point,
in $d=2$, 
\begin{eqnarray} 
z-2\zeta=0,
z-\zeta-2\chi+1-d=z-\zeta-2\chi-1=0. \nonumber\\
\label{harm exp 3}    
\end{eqnarray}

There are no graphical corrections 
$\lambda_1$ either, because the equation of motion (\ref{vperpEOM'}) 
remains unchanged by 
the transformation:
$\vec{r}_\perp \to \vec{r}_\perp-\lambda_1 \vec{v}_1 t~~~,~~
\vec{v}_\perp \to \vec{v}_\perp + \vec{v}_1~~~
$ for arbitrary constant vector $\vec{v}_1\perp\hat{x}_\parallel$. 
This  exact symmetry  must continue to hold upon renormalization, with the {\it same} value of  $\lambda_1$. Hence, $\lambda_1$ cannot be graphically renormalized.
Requiring that $\lambda_1'=\lambda_1$ in 
(\ref{rescale lambda}),
and  setting ${\rm 
graphs} =0$, implies 
$\chi =1 -z$ in {\it all} $d\le4$.
This and (\ref{harm exp 3}) forms  three independent equations for the three unknowns
$\chi$, $z$, and $\zeta$,  whose solution in $d=2$ is (\ref{exact exp3}).
The scaling exponents $z$, $\zeta$, and $\chi$ determine the scaling form of the velocity-velocity autocorrelation function in arbitrary dimension $d$ through the scaling relation\cite{TT}:
\begin{eqnarray} 
&&C_v(\vec r,t) \equiv  \langle \vec{v}_\perp(\vec 0, 0)  \cdot \vec{v}_\perp(\vec r, t)\rangle
=|\vec{r}_{_\perp} |^{2\chi} G\left({r_\parallel'\over |\vec{r}_{_\perp} |^{\zeta}}, {t\over |\vec{r}_{_\perp} |^{z}}\right)  
\nonumber\\&&
=
|\vec{r}_{_\perp} |^{2\chi} G\left({r_\parallel-\gamma t\over |\vec{r}_{_\perp} |^{\zeta}}, {t\over |\vec{r}_{_\perp} |^{z}}\right) ~,
\label{Cvscale} 
\end{eqnarray} 
where the second equality follows from scaling arguments applied to the boosted equation of motion (\ref{vperpEOM'}), and the third arises from undoing the boost.
Here $G(u, w)$ is a scaling function, with scaling arguments $u\equiv{r_\parallel-\gamma t\over |\vec{r}_{_\perp} |^{\zeta}}$ and $w\equiv{t\over |\vec{r}_{_\perp} |^{z}}$.  The asymptotic limits of $G(u,w)$
and
$C_v(\vec r,t)$  can be obtained by the following arguments.

When $r_\parallel-\gamma t\rightarrow0$ and $t\rightarrow0$, $C_v(\vec r,t)$ must clearly  depend only on $r_\perp$, and should not vanish. Hence 
$G(u\ll1,w\ll1)\rightarrow {\rm 
constant\ne0} $. This in turn implies that 
$C_v (\vec r, t) \propto |\vec r_{_\perp}|^{2\chi} $ for $ |\vec r_{_\perp}|^{\zeta}\gg|r_\parallel-\gamma t| ~,~  t^{\zeta\over z}$.
Similarly, if $\vec{r}\rightarrow0$ and $t\rightarrow0$, then $C_v(\vec r,t)$ should depend only on $|r_\parallel-\gamma t|$. This implies $G(u,w)\propto u^{{2\chi\over\zeta}}$ for $u\gg w, 1$, in order to cancel off the $|\vec r_{_\perp}|^{2\chi}$ prefactor in
Equation (\ref{Cvscale}). This in turn implies that
$C_v (\vec r, t) \propto |r_\parallel-\gamma t|^{{2\chi\over\zeta}} $ for $ |r_\parallel-\gamma t| \gg|\vec r_{_\perp}|^{\zeta}~,~  t^{\zeta\over z}$.
Similar reasoning  implies that $C_v (\vec r, t) \propto |t|^{{2\chi\over z}} $ for $ t\gg |\vec r_{_\perp}|^{z} ~,~ |r_\parallel-\gamma t|^{z\over \zeta}$.
Hence,
using the exact exponents (\ref{exact exp3}) in $d=2$, 
\begin{eqnarray}
 &&C_v (\vec r, t) 
\propto\left\{\begin{array}{ll}
 ~~~~r_\perp^{-{2\over 5}}~~~,~~~~ |r_\perp|^{3\over 5}\gg|r_\parallel-\gamma t| ~,~  t^{1\over 2}\\
 (r_\parallel-\gamma t)^{-{2\over 3}},~~|r_\parallel-\gamma t| \gg |x|^{3\over 5}~,~  t^{1\over 2} \\
 ~~~t^{-{1\over 3}}~~~,~~~~~|t| \gg|r_\perp|^{6\over 5}~,~|r_\parallel-\gamma t|^2~.
 \end{array}
 \right.
 \label{Cvscale d=2} 
\end{eqnarray}
This correlation function can be measured directly  in both simulations\cite {TT,Chate1}, and experiments \cite{Andrea and Irene}.

The relation  (\ref{rho-v}) 
between density and velocity  implies that density correlations should obey the same sort of scaling law, but with an additional power of $|r_\perp|^{-1}$ for every power of $\delta\rho$; hence, in $d=2$:
\begin{eqnarray}
 &&C_\rho (\vec r, t)  \equiv  |r_\perp|^{-{12\over 5}} G_\rho \left({r_\parallel-\gamma t\over 
|r_\perp|^{3\over 5}}, {t\over |r_\perp|^{6\over 5}}\right) 
\nonumber\\&&\propto\left\{\begin{array}{ll}
 ~~~~|r_\perp|^{-{12\over 5}}  ~~~,~~~ |r_\perp|^{3\over 5}\gg|r_\parallel-\gamma t| ~,~  t^{1\over2}\\
 (r_\parallel-\gamma t)^{-4},~~ ~~~|r_\parallel-\gamma t| \gg |r_\perp|^{3\over 5}~,~  t^{1\over2} \\
 ~~~t^{-2}~~~,~~~~~~~~|t| \gg |r_\perp|^{6\over 5}~,~|r_\parallel-\gamma t|^2~~.
 \end{array}
 \right.
 \label{Crhoscale lim} 
\end{eqnarray}
The last line holds in $d=3$ as well, because  $\chi=1-z$, does.
The last two lines of (\ref{Crhoscale lim}) directly imply  equations (\ref{persist}) and (\ref{sweet spot}).
It can also be shown \cite{me recent} that at equal times $C_\rho (\vec r, t=0)$
decays sufficiently rapidly that
there are no giant number 
fluctuations in Malthusian flocks.

I thank the Institut Poincare, the ESPCI,  
the Universite Pierre et Marie Curie, the  KITP (UCSB), CUNY,  the Lorentz Center, 
University of Leiden, and the MPI-PKS, Dresden,  for their hospitality; 
S. Ramaswamy, H. Chate, A. Cavagna, I. Giardina, M. Rao, Y. Tu, and F. Ginelli for valuable discussions;  
and K. Toner for a careful reading of the manuscript.




\begin{thebibliography}{99}
\bibitem{P2} The $P_2$ term was left out by  all of reference \cite{TT}, as was the density dependence of $\lambda_1$. The 
latter oversight 
lead to the erroneous conclusion that one could determine exact exponents for immortal flocks in $d=2$.
\end{thebibliography}
\end{document}